\documentclass[aps,prb,twocolumn,superscriptaddress,showpacs]{revtex4-1}
\usepackage{graphics}
\usepackage{graphicx}
\usepackage[english]{babel}
\usepackage{color}

\begin{document}

\title{Latent-heat and non-linear vortex liquid at the vicinity of the first-order phase transition in layered high-$T_{\rm c}$ superconductors}

\author{M.I. Dolz$^*$}
\affiliation{Departamento de F\'{i}sica, Universidad Nacional de
San Luis, and Instituto de F\'{i}sica Aplicada, CONICET, 5700 San
Luis, Argentina}
\affiliation{Laboratoire des Solides Irradi\'{e}s, Ecole
Polytechnique, CNRS URA-1380, 91128 Palaiseau, France}

\author{Y. Fasano}
\affiliation{Low Temperatures Division, Centro At\'{o}mico
Bariloche, CNEA, 8400 Bariloche, Argentina}

\author{H. Pastoriza}
\affiliation{Low Temperatures Division, Centro At\'{o}mico
Bariloche, CNEA, 8400 Bariloche, Argentina}

\author{V. Mosser}
\affiliation{Itron SAS, F-92448 Issy-les-Moulineaux, France}

\author{M. Li}
\affiliation{Kamerlingh Onnes Laboratorium, Rijksuniversiteit
Leiden, 2300 RA Leiden, The Netherlands}

\author{M. Konczykowski}
\affiliation{Laboratoire des Solides Irradi\'{e}s, Ecole
Polytechnique, CNRS URA-1380, 91128 Palaiseau, France}

\date{\today}

\begin{abstract}

In this work we revisit the vortex matter phase diagram in layered
superconductors solving still open questions by  means of AC and
DC local magnetic measurements in the paradigmatic
Bi$_{2}$Sr$_{2}$CaCu$_{2}$O$_{8}$ compound. We show that measuring
with AC magnetic techniques is mandatory in order to probe the
bulk response of vortex matter, particularly at high-temperatures
where surface barriers for vortex entrance dominate. From the
$T_{\rm FOT}$-evolution of the enthalpy and latent-heat at the
transition we find that, contrary to previous reports, the nature
of the dominant interlayer coupling is electromagnetic in the
whole temperature range. By studying the dynamic properties of the
phase located at  $T \gtrsim T_{\rm  FOT}$, we reveal the spanning in a
considerable fraction of the phase diagram of a non-linear vortex
phase suggesting bulk pinning might play a role even in the liquid
vortex phase.

\end{abstract}

 \keywords{}

\maketitle

\section{Introduction}

One of the main features of the phase diagram of vortex matter in
layered high-temperature superconductors is the occurrence of  a
first-order transition \cite{Pastoriza94a,Zeldov95a} at $T_{\rm FOT}$
due to the relevance of thermal fluctuations at temperatures close
to the critical, $T_{\rm c}$. The layered nature of vortex matter
in these extremely anisotropic materials plays a key role on the
location of $T_{\rm  FOT}$ and on the temperature-evolution of the main
thermodynamic magnitudes at the transition. When applying a field
along the sample $c$-axis vortices are actually a stack of pancake
vortices lying in CuO planes, coupling between layers via
electromagnetic and Josephson interactions \cite{Blatter}. In the
case of the paradigmatic  layered
Bi$_{2}$Sr$_{2}$CaCu$_{2}$O$_{8}$ compound the first-order
transition separates a solid phase at low temperatures and a
liquid \cite{Nelson1988} or decoupled gas \cite{Glazman1991} of
pancake vortices with reduced shear viscosity \cite{Pastoriza1995}
at high temperatures.

The solid phase presents irreversible magnetic behavior ascribed
to bulk pinning and surface barriers, each of them dominating at
different temperature and measuring-time ranges
\cite{Chikumoto1991,Chikumoto1992,Zeldov1995b}. Direct imaging of
vortices in pristine Bi$_{2}$Sr$_{2}$CaCu$_{2}$O$_{8}$ reveals the
vortex solid has quasi long-range positional order
\cite{Fasano2005,Fasano2008}.  Josephson-plasma-resonance
measurements indicate the $T_{\rm  FOT}$ transition line corresponds to
a single-vortex decoupling process between pancake vortices from
adjacent layers within the same stack \cite{Colson2003}. The
first-order line is then a single-vortex transition that depends,
at best, on the density of the surrounding vortex matter.
Therefore the relative importance of the two types of interaction
between pancake vortices determines the temperature-evolution of
the enthalpy and latent-heat, $\Delta s  T_{\rm FOT}$ of the
transition.

The main thermodynamic magnitudes entailed in this transition,
namely the entropy-jump per pancake vortex, $\Delta s$, and the
enthalpy that is proportional to the observed jump in local
induction, $\Delta B$, have been investigated experimentally as
well as theoretically \cite{Zeldov95a,Dodgson1998}. A previous
report claims that for $T_{\rm  FOT} \sim T_{\rm  c}$ the latent-heat of the
transition is not satisfactorily described by considering the
electromagnetic interaction as the dominant coupling mechanism
between pancakes of adjacent layers \cite{Dodgson1998}. The same
work proposed then the existence of a crossover to a high-
$T_{\rm  FOT}$ (low-field) regime where interlayer coupling is
dominated by the Josephson interaction.


Indeed, at such high temperatures the magnetic response of
Bi$_{2}$Sr$_{2}$CaCu$_{2}$O$_{8}$ vortex matter is dominated by
geometrical or Bean-Livingston surface barriers
\cite{Chikumoto1992,Morozov1996b}. Nevertheless, a previous work
reported on the plausibility of bulk pinning playing a role in
vortex dynamics even at temperatures larger than $T_{\rm  FOT}$, within
the vortex liquid phase \cite{Indenbom1996}. Bulk pinning do play
a dominant role for temperatures below $0.6 T_{\rm c}$, and the
first-order transition is observed as the so-called second-peak
effect \cite{Avraham2001,Chikumoto1991} or order-disorder
transition, $H_{\rm  SP}$ \cite{Khaykovich97a,Vinokur1998,Cubbit1993}.
This region of the transition is detected through an increase
of the width of DC hysteresis loops \cite{Chikumoto1991}, or as a
minimum in AC magnetization loops, both occurring at $H_{\rm SP}$,
independently of frequency.

Therefore several issues in the vortex phase diagram of layered
high-temperature superconductors remain still open to discussion.
Most of the attempts in order to settle down this debate suffered
from comparing data obtained with experimental techniques lacking
on the proper sensibility in order to ascertain the relative
importance of bulk effects, surface barriers, and sample
inhomogeneities. In this work, we apply AC
local-magnetic-measurement techniques with the aim of providing an
accurate description of the physics entailed in the first-order
vortex phase transition in Bi$_{2}$Sr$_{2}$CaCu$_{2}$O$_{8}$. We
rule out the proposal of a crossover between electromagnetic to
Josephson coupling taking place at low fields (high-temperatures),
suggesting the dominant interaction between pancakes at $T_{\rm  FOT}$
is always of the same nature. In addition, we report on a
non-linear liquid vortex phase spanning a considerable fraction of
the high-temperature region of the  phase diagram. The large
extent of this phase challenges the idea of a vortex liquid being
a linear phase.

\begin{figure}[ttt]
\includegraphics[width=\columnwidth,angle=0]{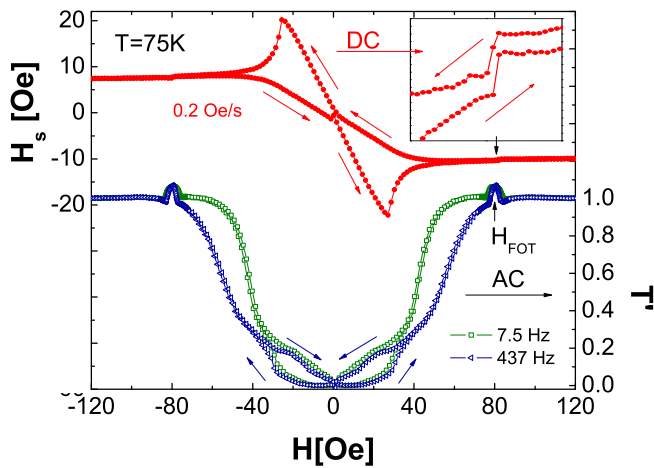}
\caption{DC and AC magnetic hysteresis loops of Bi$_{2}$Sr$_{2}$CaCu$_{2}$O$_{8}$ sample measured at 75\,K.
Top: ascending and descending branches of a DC hysteresis loop.
Insert: zoom-in of the DC  loop at the vicinity of the first-order
transition entailing a $B$-jump (see black arrow). Red arrows
indicate the ascending and descending branches. Bottom: AC
transmittivity loops with paramagnetic peaks fingerprinting the
first-order transition field $H_{\rm FOT}$. The loops were
measured with ripple fields of 0.9\,Oe rms in amplitude and
frequencies of 7.5 and 437\,Hz. Blue arrows indicate the ascending
and descending branches. \label{figure1}}
\end{figure}

\section{Experimental}

The optimally-doped Bi$_{2}$Sr$_{2}$CaCu$_{2}$O$_{8}$
single-crystal studied in this work ($T_{\rm c}=90$\,K) was grown
by means of the traveling-solvent floating zone technique
\cite{Li94a}. The local magnetization of the $220 \times 220
\times 30$\,$\mu$m$^3$ sample was measured with a microfabricated
2D electron gas Hall-sensor array \cite{Konczykowski91a}. The
eleven-probes array was photolithographically fabricated from
GaAs/AlGaAs heterostructures and having each sensor an active area of $6 \times
6$\,$\mu$m$^{2}$.

\begin{figure} [ttt]
\includegraphics[width=\columnwidth,angle=0]{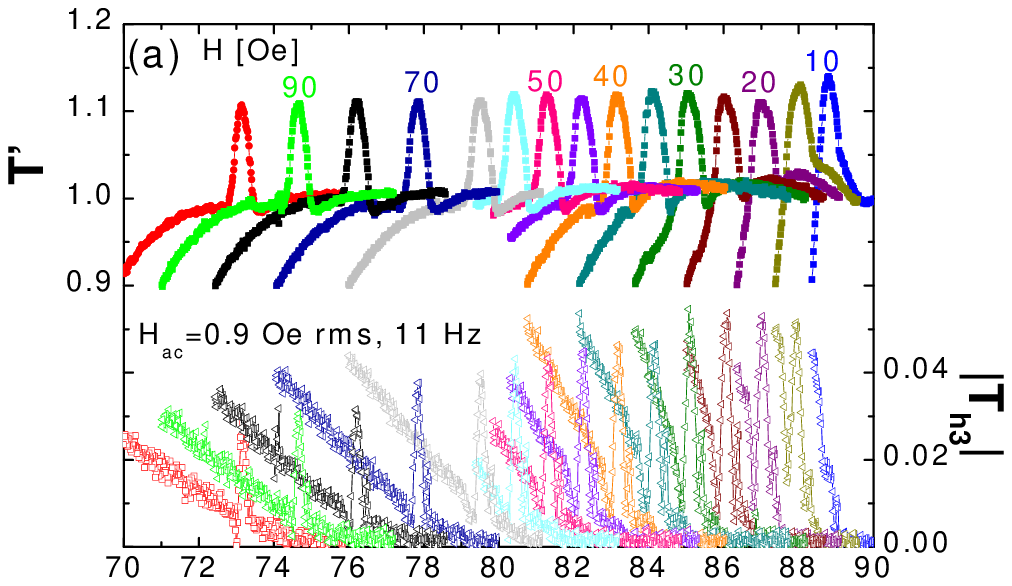}
\includegraphics[width=\columnwidth,angle=0]{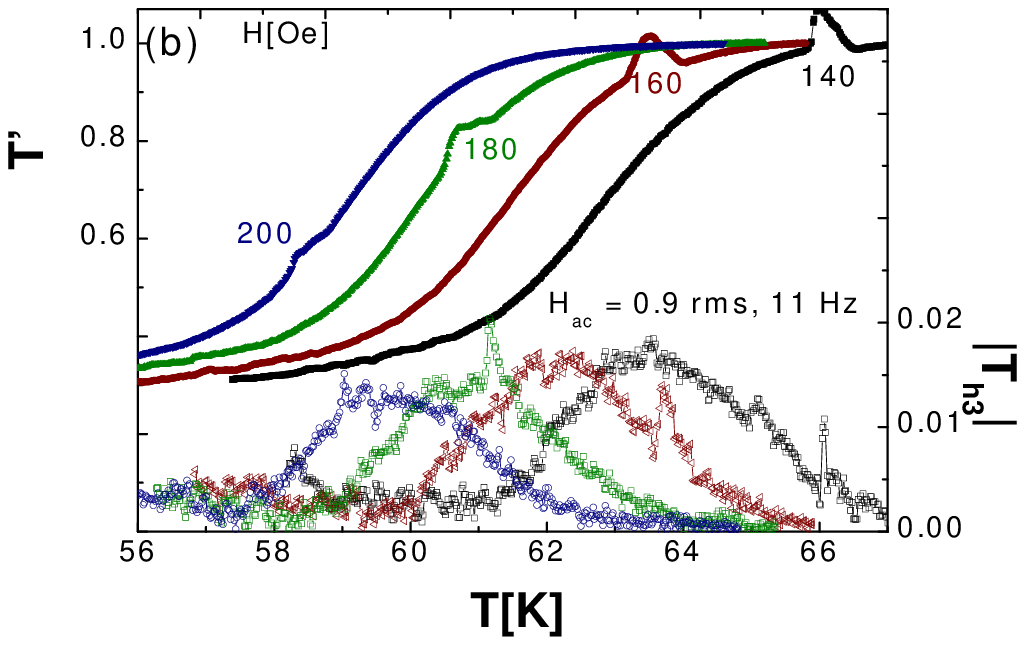}
\caption{Temperature-dependence of the transmittivity and modulus
of the third harmonic response for
Bi$_{2}$Sr$_{2}$CaCu$_{2}$O$_{8}$ vortex matter. (a) Very-low and
(b) low field regimes. The AC ripple field of 0.9\,Oe rms and
11\,Hz is collinear to the applied field $H$. \label{figure2}}
\end{figure}

Local AC magnetization measurements were performed applying  DC
and ripple magnetic fields parallel to the $c$-axis, $H$ and
$H_{\rm ac}$ respectively. DC magnetic hysteresis loops are
obtained by measuring the magnetization, $H_{\rm s} = B - H$, when
cycling $H$ at fixed temperatures. In the AC measurements the
ripple field has an amplitude of $0.9$\,Oe rms and frequencies
ranging from $1$ to $1000$\,Hz. The first and third-harmonics of
the AC magnetic induction are simultaneously measured by means of
a digital-signal-processing lock-in technique. The in-phase
component of the first-harmonic signal, $B'$, is converted to the
transmittivity \cite{Gilchrist1993}, $T'$, a magnitude extremely
sensitive to discontinuities in the local induction associated to
first-order magnetic transitions. A non-negligible magnitude of
the third harmonic signal, $\mid T_{\rm h3} \mid$, indicates the
appearance of non-linearities in the magnetic response. The
onset of $\mid T_{\rm h3} \mid$ on cooling is considered as
the irreversibility temperature or field, $T_{\rm  IL}$ or $H_{\rm  IL}$
\cite{vanderBeek1995}.

In order to track the $H\,-\,T$ location of the first-order
transition and irreversibility lines in
Bi$_{2}$Sr$_{2}$CaCu$_{2}$O$_{8}$ vortex matter, we perform two
types of measurements: Isothermal DC and AC hysteresis loops
\cite{Konczykowski2006}, and temperature-evolution of $T'$ and
$\mid T_{\rm h3} \mid$ on field-cooling at various magnetic fields
(see Figs.\,\ref{figure1} and \ref{figure2}).  We discuss the
magnetic response of our sample at three characteristic
temperature regimes of the vortex phase diagram.

\section{Results and discussion}

In the high-temperature regime, $T \gtrsim 0.83\, T_{\rm c}$, the
first-order transition is manifested in $T'$ as a prominent
paramagnetic peak developing at the same $H_{\rm FOT}$ than the
jump in $B$ detected in DC hysteresis
loops\cite{Morozov1996,Konczykowski2006}. The top panel of
Fig.\,\ref{figure1} shows a typical two-quadrant DC loop observed
in this temperature regime indicating surface barriers dominate
the vortex entrance to the sample. Closer inspection of the DC
data in the vicinity of 80\,Oe reveals a $H_{\rm s}$ jump with
similar height for both, ascending and descending branches. This
feature is the fingerprint of the first-order transition $H_{\rm
FOT}$ in the high-temperature region \cite{Zeldov95a}. The bottom panel of
Fig.\,\ref{figure1} shows that in the AC loops this transition is detected
with improved resolution: Paramagnetic peaks appear at the same fields where the $H_{\rm s}$
jumps are measured \cite{Konczykowski2006}. At lower fields the shielding capability is increased.
On increasing frequency, the system enhances its shielding capability
manifested as a $T'$ decrease. The field-location of the
paramagnetic peak in AC loops, $H_{\rm FOT}$, is
frequency-independent.

Figure \ref{figure2} (a) shows a set of AC magnetic data for
applied fields up to 100\,Oe. The paramagnetic peak, observed in
$T'$ vs. $T$ measurements at $T_{\rm FOT}$, shifts towards lower
temperatures on increasing field. The peaks are sharp with an
amplitude that slightly decreases on increasing field. Figure
\ref{figure2} (b) shows that at fields larger than 100\,Oe shielding
currents develop in the sample prior to the appearance of the
paramagnetic peak. Nevertheless, the peak can be clearly detected
in AC magnetic measurements up to 200\,Oe.

\begin{figure} [ttt]
\includegraphics[width=0.9\columnwidth,angle=0]{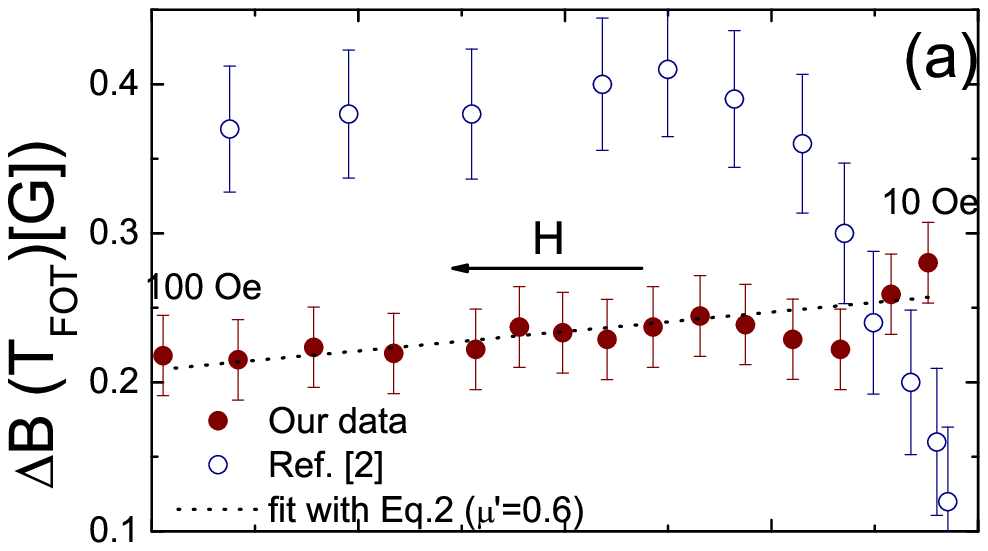}
\includegraphics[width=0.9\columnwidth,angle=0]{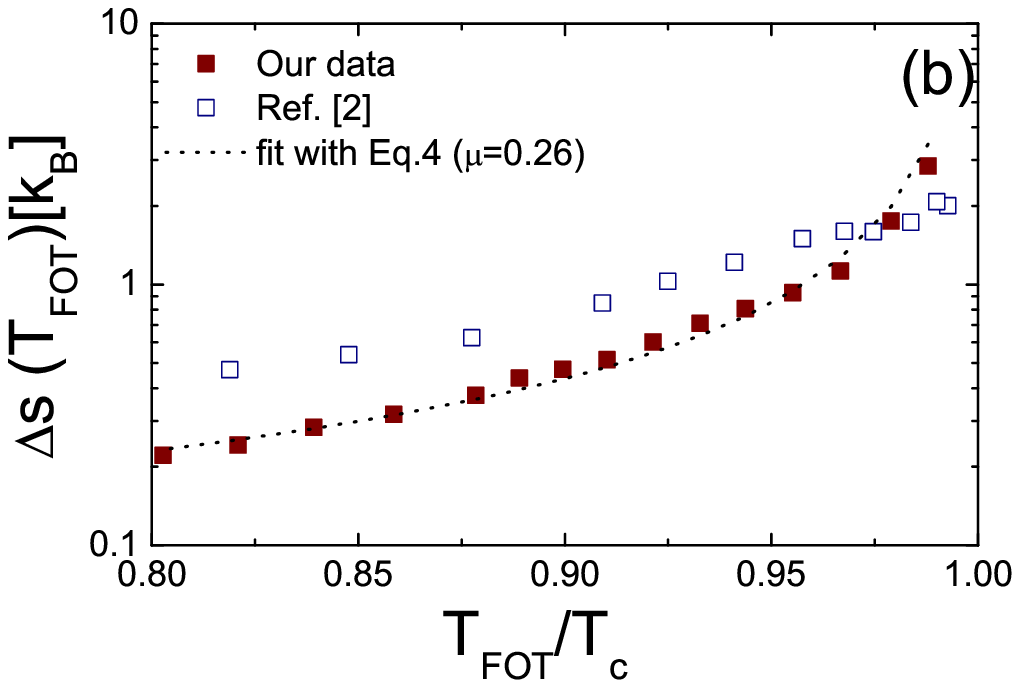}
\caption{(a) Magnitude of the magnetic-induction jump, $\Delta B$,
and (b) of the entropy jump per pancake vortex, $\Delta s$, at the
first-order transition of Bi$_{2}$Sr$_{2}$CaCu$_{2}$O$_{8}$ vortex
matter. Our $\Delta B$ and $\Delta s$ data are reasonably well fitted assuming dominant electromagnetic interaction between pancake
vortices in the whole temperature range (see Eqs.\,\ref{eqDB} and
\ref{eqDstheoric} in the text).  Data from Ref.\,\onlinecite{Zeldov95a} are plotted for comparison.
\label{figure3}}
\end{figure}

The enthalpy of this first order transition is proportional to the
height of the step in the magnetic induction at $T_{\rm FOT}$. This
magnitude can be obtained from the transmittivity value at the
transition field considering that $T' = B'/H_{\rm  ac}$ and according to Ref.\,\onlinecite{Morozov1996}

\begin{eqnarray}
T'= 1+\frac{2 \Delta B}{\pi H_{\rm ac}} \label{eqDB}
\end{eqnarray}

\noindent for small ripple fields $H_{\rm ac}$. Figure \ref{figure3}
(a) shows the $\Delta B$ evolution with $T_{\rm FOT}/T_{\rm c}$
obtained from applying Eq.\,\ref{eqDB} to $T'(T)$ data at
different fields. We found a linear increase of $\Delta B$ up to
temperatures $T_{FOT} \sim  T_{c}$ within the error bars. These
findings are in contrast to the seminal work of
Ref.\,\onlinecite{Zeldov95a} reporting that for $T_{FOT} \geq 0.93
T_{c}$ the jump in magnetic induction decreases dramatically down
to 25\% at $T_{c}$ (see Fig.\,\ref{figure3} (a)). This result was
interpreted within the framework of a crossover between an
electromagnetic to Josephson coupling of pancake vortices on
decreasing field (increasing $T_{FOT}$). This interpretation was
proposed by the theoretical work of Ref.\,
\onlinecite{Dodgson1998} that provided the functionality of
$\Delta B (T)$ at the first-order transition.

A linear evolution of $\Delta B$ can be explained by considering
that the inter-layer coupling is dominated by electromagnetic
interactions between weakly bound pancake vortices undergoing
large thermal fluctuations. Taking into account the dispersive
electromagnetic line tension of vortices and the Lindemann
criterion for vortex melting, \cite{Dodgson1998,Blatter} in this case

\begin{eqnarray}
\Delta B= \mu ' \frac{k_{\rm B} T_{\rm FOT}}{\phi_{\rm 0} d}  \label{eqDstheoric}
\end{eqnarray}

\noindent where $k_{\rm B}$ is the Boltzmann constant, $\phi_{\rm
0}$ the flux quantum, $d \approx 15 \AA$ the CuO planes
inter-layer distance and $\mu '$ a numerical constant
\cite{Dodgson1998}. The latter constant depends on the Lindemann number
for a given material and up to our knowledge
there is no report of its quantitative value in the case of
Bi$_{2}$Sr$_{2}$CaCu$_{2}$O$_{8}$. The linear $\Delta B$ data
obtained in our sample is reasonably well fitted with
Eq.\,\ref{eqDstheoric} for $\mu '=0.6$.

The entropy-jump per vortex and per CuO layer entailed in the
first-order transition can be obtained from $\Delta B$ data by
means of the thermodynamic Clausius-Clapeyron relation

\begin{eqnarray}
\Delta s= -\frac{\phi_{\rm 0} d}{4\pi}\frac{\Delta B}{B_{\rm FOT}} \,
\frac{dH_{\rm FOT}}{dT}. \label{eqDs}
\end{eqnarray}

\noindent Figure \ref{figure3} (b) shows that the $\Delta s$ data
of our sample diverges close to $T_{c}$, as also reported in the
previous work of Ref.\,\onlinecite{Zeldov95a}. In order to explain
the measured $T_{FOT}$ evolution of $\Delta s$ in terms of the
nature of the inter-layer coupling that dominates in the system,
Ref.\,\onlinecite{Dodgson1998} proposes that for electromagnetic
interactions,

\begin{eqnarray}
\Delta s= \frac{\mu}{\pi} \frac{k_{\rm B}}{1-(T_{\rm FOT}/T_{\rm
c})} d  \label{eqDs2}
\end{eqnarray}

\noindent with $\mu$ a material-dependent constant. Our $\Delta s$
data is reasonably well fitted with the expression of
Eq.\,\ref{eqDs2} in the whole temperature range close to $T_{c}$.
Therefore, our results are in discrepancy with the data and
interpretation provided in Ref.\,\onlinecite{Zeldov95a} where the divergent $\Delta s$ was not satisfactorily
fitted with Eq.\,\ref{eqDs2}. However this might be associated to
the fact that in Ref.\,\onlinecite{Zeldov95a} $\Delta B$ decreases
close to $T_{c}$. The difference between our data and those of
Ref.\,\onlinecite{Zeldov95a} can have origin in our AC
measurements being more sensitive to bulk currents than the DC
measurements reported in Ref.\,\onlinecite{Zeldov95a}, more
affected by surface-barrier effects. However, inhomogeneities in the seminal sample
can not be ruled out as possible sources of this discrepancy.

Another remarkable phenomenology in the high-temperature region of
the phase diagram is that the irreversibility temperature, $T_{\rm
IL}$, identified from the onset of the third-harmonic signal on
cooling, develops at temperatures larger than those associated to
the paramagnetic peak (see the low panel of Fig.\,\ref{figure2}
(b)). This indicates the existence of a phase region with
non-linear vortex dynamics at fields exceeding $H_{\rm FOT}$. A
previous work reported a qualitatively similar phenomenology by
means of magneto-optics techniques and proposed that bulk pinning
still plays a role in the high-temperature liquid vortex phase
\cite{Indenbom1996}. Here we study in detail the quantitative
spanning of this non-linear vortex liquid  as a function of
frequency.

\begin{figure} [ttt]
\includegraphics[width=\columnwidth,angle=0]{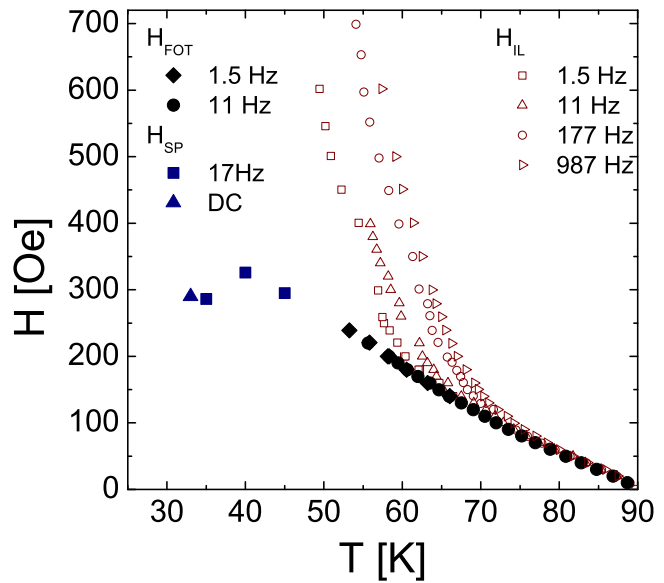}
\caption{Vortex phase-diagram of Bi$_{2}$Sr$_{2}$CaCu$_{2}$O$_{8}$
depicting the first-order transition line in the high and low
temperature regions, $H_{FOT}$ and $H_{SP}$ (full points). Data
were obtained from AC and DC magnetic measurements. The
frequency-dependent irreversibility line, $H_{IL}$, is indicated
with open points. All AC measurements were performed applying a
ripple field with amplitude of 0.9\,Oe rms.} \label{figure8}
\end{figure}

Figure \ref{figure8} shows that as higher applied fields the
shielding of the AC field starts at significantly higher
temperatures than the occurrence of the first-order transition.
The temperature of this phase increases with
frequency and at high fields (300\,Oe) can be as large as 20\% of
$T_{c}$ for frequencies of the order of 1\,kHz. This phenomenon
might have origin from a residual effect of pinning
\cite{Vinokur1991}, or Bean-Livingston barriers \cite{Fuchs1998}
on the high-temperature liquid phase.

\begin{figure}
\includegraphics[width=\columnwidth,angle=0]{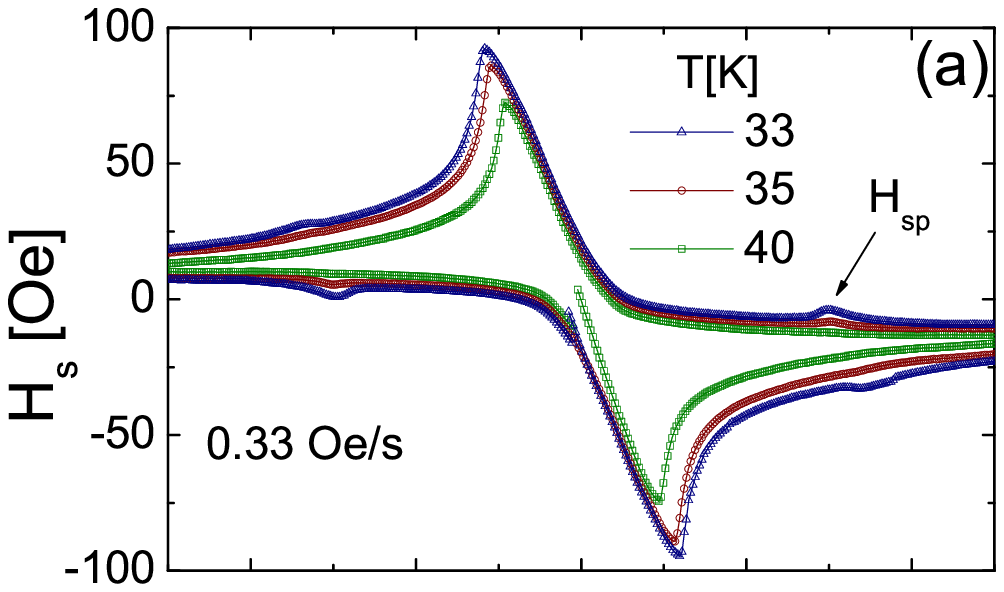}
\includegraphics[width=\columnwidth,angle=0]{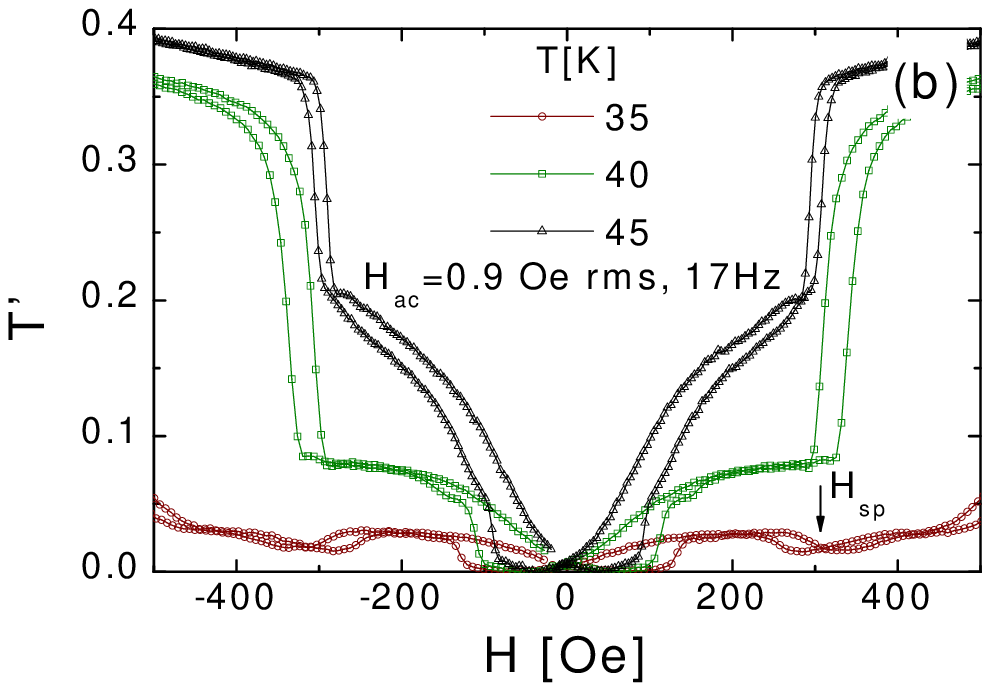}
\caption{DC and AC magnetic hysteresis loops for
Bi$_{2}$Sr$_{2}$CaCu$_{2}$O$_{8}$ vortex matter in the
low-intermediate temperature regime. (a) DC loops: the transition
field $H_{\rm SP}$ is taken at the midpoint between the onset and
the full development of the local minimum in the ascending branch.
(b) Transmittivity AC loop measured with a ripple field of 0.9\,Oe
rms and 17\,Hz parallel to $H$. The $H_{\rm SP}$  field is
obtained similarly as in DC loops (see arrow). The field location
of the step-like feature, $H_{\rm step}$, is taken at half the
step height in the ascending field branch. \label{figure5}}
\end{figure}

The vortex phase diagram of Fig.\,\ref{figure8} also shows the
connection between the low-temperature regime of the first-order
transition, $H_{SP}$, and the high-temperature regime $H_{FOT}$
for $0.4 \lesssim  T/T_{c} \lesssim 0.5$ \cite{}. The $H_{\rm SP}$ transition field can be obtained from DC magnetic
hysteresis loops as shown in Fig.\,\ref{figure5}(a) for three
different temperatures. In these curves the local minimum
(maximum) of the ascending (descending) branch becomes more
evident on cooling. The loops occupy two field-quadrants and
noticeably increase their width on cooling. This figure also
indicates that the $H_{\rm SP}$ field is hard to detect from DC
magnetic loops for temperatures larger than 33\,K. This is due to
a technical limitation of the DC technique that lacks resolution
in order to measure bulk transitions when surface barriers
dominate the magnetic properties \cite{Chikumoto1992}.

We therefore use the AC hysteresis loop technique in order to
track the second-peak transition in the intermediate
temperature-regime. The AC transmittivity reflects the
dimensionless normalized sustainable-current density,
$J=j(f)a/H_{\rm ac}$, from $J = (1/\pi)\arccos(2 T' - 1)$
\cite{vanderBeek1995}. This formula was derived for an AC
penetration in the Bean critical regime, an assumption that seems
to be valid in view of the results presented in
Ref.\,\onlinecite{Indenbom1996}. A low-temperature AC loop at
35\,K is shown in Fig.\,\ref{figure5} (b) depicting local minima
in both, the ascending and descending branches. This minima can be
ascribed to the second-peak transition $H_{\rm SP}$ since a
minimum in $T '$ corresponds to a maximum in the bulk $J$. The
$T'$ signal evolves in a different manner for the high- and
intermediate-temperature regimes. For temperatures larger than
$0.66\,T_{\rm c} =58$\,K, the transmittivity presents paramagnetic
peaks, developing at the flanks of the central depletion (see
Fig.\,\ref{figure2}). For intermediate temperatures a sudden jump
of $T'$ is detected, as for instance in the AC loops measured at
40 and 45\,K (see Fig.\,\ref{figure5} (b)). This step-like
feature, manifested at a field $H_{\rm step}$ implies a
frequency-independent  drop of the magnetic hysteresis, consistent
with the change in the width of the DC loops, and indicating that
this transition is governed by the local value of $B$ rather than
$H$ \cite{Khaykovich1996}.  The jump in $T'$ is related to the
sudden change in shielding currents with the sample becoming more
transparent to the penetration of the AC ripple field at larger
$H$. Similarly, the height of the steps in $T '$ decreases on
increasing temperature.

\begin{figure}
\includegraphics[width=\columnwidth,angle=0]{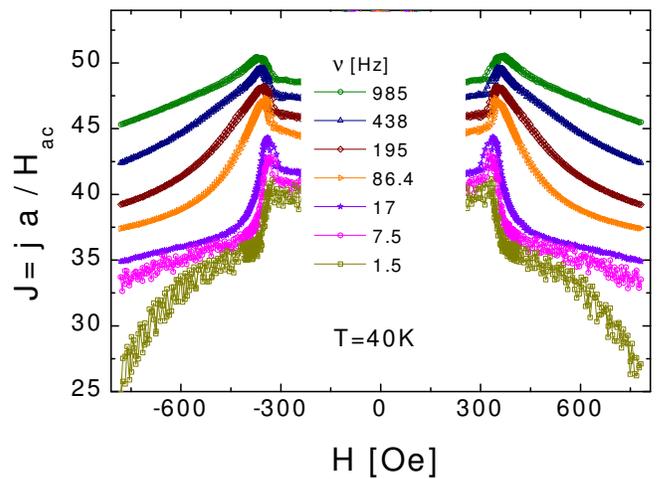}
\caption{Normalized sustainable-current density $J$ as a function
of applied field at 40\,K. The
frequency dependence of $J$ is obtained from AC magnetization
loops measured with a ripple field of 0.9\,Oe rms and frequencies
ranging 1.5 to 985\,Hz. \label{figure6}}
\end{figure}

The crossover temperature for the detection of the second-peak
\cite{Sochnikov2008,vanderBeek1996}, or of the step-like feature,
is time/frequency dependent \cite{Chikumoto1992}. Figure\,\ref{figure6} presents the
evolution of the normalized sustainable-current density extracted
from $T'$ as a function of frequency in the range from 1.5 to
985\,Hz. For frequencies smaller than 7.5\,Hz a sudden drop of $J$
develops at fields $ H_{\rm SP} \sim 330$\,Oe. On increasing
frequency, this feature evolves into a field-asymmetric peak
producing the step-like feature observed in $T'$ curves. The onset
of this peak, or equivalently the $H_{\rm step}$ field in $T'$, is
frequency independent.

Thus, for $0.39\,T_{\rm c}\approx 35K < T < 0.66\,T_{\rm c}\approx
60K$ we detect, at a field $H_{\rm step}$, a discontinuous
decrease of the bulk shielding-currents associated with the
first-order transition. Below $T=0.39\,T_{\rm
c}=35$\,K the opposite effect an increase of the shielding
currents is observed at almost the same field indicating the
$H_{\rm SP}$ transition. In the vicinity of this reversal of
current behavior, varying the frequency of the AC ripple field
tunes a decrease (low-frequencies) or an increase
(high-frequencies) of the shielding currents. The latter case is
equivalent to probing magnetic relaxation on a shorter time-scale,
in analogy to DC magnetization experiments \cite{Chikumoto1992},
or to choosing a higher electric field in a transport $I(V)$
measurement. Since the $I(V)$ curves just below and above the
first order transition cross --- with the electric field in the
high-field phase being larger than that in the low-field phase for
the low-current density limit, and vice-versa for the high-current
density limit ---, varying the working point (by tuning the
frequency) leads to either a step like-behavior of the screening
current (at low electric fields) or a peak-like curve (at high
electric fields) \cite{Konczykowski2000}. The energy barriers for
flux creep have $U(J)$ and $E(J)$ curves with different
functionalities for fields larger or smaller than the transition
one. On varying field close to the transition these curves cross,
and the phase transition produces a discontinuous change on the
electrodynamics of vortex matter. Detecting the transition with a
high electric field (short measurement times) leads to an
enhancement of shielding currents with increasing field, whereas
with a low electric field (long times) a sudden decrease is
observed.

\section{Conclusions}

We revisited the vortex phase diagram in pristine
Bi$_{2}$Sr$_{2}$CaCu$_{2}$O$_{8}$ vortex matter by applying AC and
DC local magnetization techniques and reveal a new phenomenology
regarding the thermodynamic properties and the nature of the
first-order phase transition line. We show that in this case AC
local-magnetic techniques probe the physics at shorter
time-scales, having access to currents flowing in the bulk sample
with improved sensitivity than DC measurements. As a consequence,
we were able to track down the connection between the low $H_{\rm SP}$
and high $H_{\rm FOT}$ temperature regions of the first-order vortex
phase transition in an intermediate temperature-regime, covering
the whole temperature-range of the transition. By combining
simultaneous measurements of the first and third-harmonics of the
magnetization we were also able to detect the existence of a
non-linear vortex liquid phase at $T < T_{\rm IL}$ spanning a vortex
phase region that at 300\,Oe can be as large as 20\,K. The width
of this non-linear liquid region increases on increasing
frequency. From AC magnetic data we also found that, contrary to
previous reports \cite{Zeldov95a}, the enthalpy associated to the
high-temperature first-order transition, $\propto \Delta B$,
increases linearly with $T_{\rm FOT}$ in the whole temperature-range
up to $T_{\rm c}$. By means of the thermodynamic Clausius-Clapeyron
relation we estimated the entropy-jump per pancake vortex. We
explain the temperature-evolution of the latent-heat of our sample
considering solely the dominance of electromagnetic inter-layer
coupling all along the $H_{\rm FOT}$ transition. This is in clear
contrast to the claim of Ref.\,\onlinecite{Zeldov95a} that during
the $H_{\rm FOT}$ the dominant interlayer coupling changes nature on
approaching $T_{\rm c}$.

\section{Acknowlodgements}

We thank to C.J. van der Beek for selecting the crystals. This work
was supported by the ECOS Sud-MINCyT France-Argentina
collaboration program, Grant No. A09E03 and by PICT-PRH 2008-294
from the ANPCyT.

$^*$ To whom correspondence should be addresed:
mdolz@unsl.edu.ar

\end{document}